\documentclass[sigconf, 9pt, nonacm]{acmart}

\AtBeginDocument{%
  }
    
\renewcommand\footnotetextcopyrightpermission[1]{}

\usepackage[compact]{titlesec}

\usepackage{entlab}

\titlespacing*{\subsubsection}
{0pt}{.5ex}{.5ex}
\titlespacing*{\subsection}
{0pt}{.7ex}{.5ex}

\newcommand{\sysname}{{PEARL}\xspace}

\begin{document}

\title{\sysname: Power- and Energy-Aware Multicore Intermittent Computing}

\author{Khakim Akhunov}
\affiliation{%
 \institution{imec}
 \city{Leuven}
 \country{Belgium}}
\email{khakim.akhunov@imec.be}
  
\author{Eren Y{\i}ld{\i}z}
\affiliation{%
 \institution{Georgia Institute of Technology}
 \city{Atlanta}
 \country{USA}}
\email{eyildiz8@gatech.edu}

\author{Kas{\i}m Sinan Y{\i}ld{\i}r{\i}m}
\affiliation{%
  \institution{University of Trento}
  \city{Trento}
  \country{Italy}}
\email{kasimsinan.yildirim@unitn.it}

\begin{abstract}

Low-power multicore platforms are suitable for running data-intensive tasks in parallel, but they are highly inefficient for computing on intermittent power. In this work, we present \sysname (PowEr And eneRgy-
aware MuLticore Intermittent Computing),  a novel systems support that can make existing multicore microcontroller (MCU) platforms suitable for efficient intermittent computing. \sysname achieves this by leveraging only a three-threshold voltage tracking circuit and an external fast non-volatile memory, which multicore MCUs can smoothly interface. \sysname software runtime manages these components and performs energy- and power-aware adaptation of the multicore configuration to introduce minimal backup overheads and boost performance. Our evaluation shows that \sysname outperforms the state-of-the-art solutions by up to 30$\times$ and consumes up to 32$\times$ less energy. 
\end{abstract}




\date{}
\maketitle
\pagenumbering{gobble}

\pagestyle{plain}

\section{Introduction}
\label{sec:introduction}

Batteryless devices use their energy harvesters to convert radio waves, sunlight, or heat into electrical energy, which is then stored in energy storage capacitors~\cite{ahmed2024internet}. These capacitors store limited energy, enabling these devices to perform only part of the computational tasks. As a result, battery-free devices frequently experience power failures caused by depleted capacitors, leading to {\em intermittent execution}~\cite{lucia2017intermittent}. When there is a power failure, the volatile program state may be lost. Therefore, it should be backed up in non-volatile memory before a power failure and recovered after the power is restored. Developing software/hardware support for efficient backup and recovery is the main focus of current research on intermittent computing~\cite{choi2022compiler}.

Currently, on-device intelligence is becoming essential for batteryless edge applications, making artificial intelligence (AI) inference the standard computational workload for these devices~\cite{gobieski2019intelligence, farina2024memory,custode2024fast,barjami2024intermittent}. Since energy harvesting occurs at irregular intervals, and longer charging times can lead to increased latency during computations, it is essential to reduce the total time required for inference and increase throughput. This is crucial for preventing stale computations and ensuring timely responses~\cite{hester2017timely,kortbeek2020time,erata2023etap,yildiz2024adaptable}. Distributing data-intensive inference workloads across multiple cores and executing them in parallel can speed up inference on batteryless devices significantly~\cite{akhunov2022adamica}. 
However, the majority of intermittent computing studies target only single-core platforms~\cite{abubakar2022protean}, overlooking the growing interest in batteryless inference that requires parallel intermittent execution of data-intensive tasks. Low-power multicore platforms in the market are ideal for parallel execution on continuous power, but intermittent computing on these architectures is challenging~\cite{akhunov2022adamica}. We identify the following two core challenges.

{\bf Lack of fast non-volatile memory (C1).} Off-the-shelf multicore platforms include only flash memory to store program code and data memory, which is very expensive in terms of memory access latency and energy. 
This is the main reason that made  Texas Instrument's single-core MSP430FR series~\cite{msp430fram} microcontrollers (MCUs) the de facto platform for intermittent computing since these MCUs include fast and energy-efficient non-volatile memory, i.e., FRAM (ferroelectric random access memory)~\cite{FRAM}. Compared to flash memory, which requires the erasure of segments to
create writable memory space and has a very low endurance ($10^5$ erases~\cite{FLASH}) for intermittent computing, FRAM is erase-free and has an extremely high endurance ($10^{13}$ reads/writes~\cite{spifram}). Besides, FRAM read is approximately $3\times$ faster and $4\times$ cheaper concerning energy consumption, while FRAM write is approximately $7\times$ faster and $123\times$ cheaper~\cite[Table 1]{chen2024ifkvs}. Therefore, intermittent computing on multicore MCUs with only flash memory is very inefficient due to the energy cost of frequent non-volatile memory accesses and problematic due to the low endurance of flash memories.


{\bf Adaptation overhead (C2).} Harvested energy dynamics impact the charging time of the energy storage capacitor and, in turn, the throughput significantly~\cite{desai2020power,akhunov2022adamica}. When charging, batteryless devices are off and not computing. The computation can be sped up significantly by executing parallelizable workloads on multiple cores, but this might come with the price of faster depletion of the capacitor, frequent power failures, longer charging times, and increased backup and restoration overheads. Therefore, enabling the most performant multicore configuration is not always preferable to increase throughput~\cite{akhunov2022adamica,balsamo2019momentum}. Depending on the ambient power, even single-core execution of parallelizable workloads might have better throughput compared to multicore execution.

In short, multicore computing support is crucial for boosting the throughput of batteryless applications. However, current multicore platforms are designed for continuous operation, making them highly inefficient for intermittent computing. To the best of our knowledge, no prior work has provided the necessary systems support to enable efficient multicore intermittent computing on these platforms. AdaMICA~\cite{akhunov2022adamica}, the closest work to this paper, is a simulation/emulation-based work that mainly deals with C2 and strictly requires embedded internal FRAM in multicore architecture, which does not exist on the market as of now (C1). 

{\bf Contributions.} We introduce {\bf \sysname} ({\bf P}ow{\bf E}r {\bf A}nd ene{\bf R}gy-aware Mu{\bf L}ticore Intermittent Computing), a novel systems support that enables efficient intermittent computing on the common off-the-shelf low-power multicore MCU platforms. 
\sysname addresses C1 by utilizing an off-the-shelf external FRAM module~\cite{spifram} that can be easily connected to the serial peripheral interface (SPI) of multicore MCUs.   
The downside of this approach is that SPI communication is energy-hungry. Moreover, most external non-volatile memories are single-ported, leading to scalability issues for multicore architectures. To overcome this inefficiency, \sysname limits FRAM access to only backup and restore operations, allowing multicore MCUs to use SRAM as main memory for computation, hence, exploiting the efficiency of their internal memory hierarchy. 
\sysname further optimizes FRAM access by minimizing backup and restore operations through {\em energy awareness} by adopting three threshold voltage tracking~\cite{akhunov2023enabling,lukosevicius2017using} solutions to multicore MCUs. \sysname backups only when the stored energy in the capacitor drops below the backup threshold and the capacitor continues to discharge since the ambient power is smaller
than the deep sleep mode power consumption of the multicore platform. This strategy significantly reduces the frequency of backups. 
Finally, \sysname increases 
throughput via {\em power awareness} by estimating ambient power level in a cheap way and adapts throughput, considering
environmental power dynamics by switching to the most performant multicore configuration.

We evaluated \sysname via simulations and experiments in our testbed using MAX32666~\cite{max32666}, an ultra-low-power MCU featuring a dual ARM Cortex-M4F processor. 
Our evaluation shows that \sysname outperforms the state-of-the-art solutions by up to 30$\times$ and consumes up to 32$\times$ less energy. Overall, we make the following key contributions:

\begin{compactenum}
    \item {\bf Adoption of Multicore MCUs.} \sysname requires only a three-threshold voltage tracking circuit and an external SPI-based FRAM connected to existing ultra-low-power multicore MCUs to enable multicore intermittent computing on these platforms. 

    \item {\bf Adaptive Intermittent Runtime.} \sysname software runtime performs energy- and power-aware multicore adaptation, introduces minimal backup overheads, and boosts performance.
\end{compactenum}

We release \sysname as open source via~\cite{pearl-repository} for the research community, filling an important gap by providing the missing multicore intermittent computing support to foster the widespread adoption of batteryless computing.



\setlength{\tabcolsep}{0.15em}   
\newcolumntype{C}[1]{>{\centering}m{#1}}
\begin{table*}
	\centering
	\scriptsize
     \caption{Comparison of the main features of \sysname  with the prior art.}
	\begin{tabular}{
			>{\centering}m{0.08\textwidth}
			>{\centering}m{0.07\textwidth}
			>{\centering}m{0.1\textwidth}
			>{\centering}m{0.07\textwidth}
            >{\centering}m{0.16\textwidth}
            >{\centering}m{0.16\textwidth}
            >{\centering}m{0.15\textwidth}
			m{0.14\textwidth}<{\centering}
			}
		\toprule		

        \rowcolor{black!3}

		\textbf{SOTA} &
		\textbf{Multicore Support} &
		\textbf{Parallel Intermit. Programming} & 
		\textbf{Intermit. Software Support} &
		\textbf{Power Adaptation Support} & \textbf{Sleep Mode Support} & \textbf{Power Awareness} & \textbf{Platform Support} \\
			
		\midrule        
         Rehash~\cite{bakar2021rehash} & No \xmark & No \xmark & Yes \cmark & Software scaling \xmark & No \xmark & Heuristics \xmark & MSP430FR \xmark \\
        
        \rowcolor{black!3}D\textsuperscript{2}VFS~\cite{ahmed2020intermittent,maioli2025dynamic} & No \xmark & No \xmark & Yes \cmark & DVFS \xmark & No \xmark & ADC \xmark & MSP430G with SPI FRAM \xmark \\

        PowerNapping~\cite{daulby2020improving} & No \xmark & No \xmark & No \xmark & No \xmark & Yes, with NO sleep mode checkpoints \cmark & ADC \xmark & MSP430FR \xmark \\ 
  
	\rowcolor{black!3}TETRA~\cite{akhunov2023enabling} & No \xmark & No \xmark & No \xmark & No \xmark & Yes, with NO sleep mode checkpoints \cmark  & No \xmark & MCU with SPI FRAM \xmark \\

        RockClimb~\cite{choi2022compiler} & No \xmark & No  \xmark & Yes \cmark
		& No \xmark & Yes but with compiler-placed checkpoints \xmark & No \xmark & MSP430FR \xmark \\
  
		\rowcolor{black!3}Momentum~\cite{balsamo2019momentum} & Yes \cmark & No \xmark & No \xmark & Core hot-plugging \cmark & Yes, but checkpoints before entering sleep mode \xmark & External circuit with ADC \xmark & Multicore SoC with Flash \xmark \\ 
        
        DVFS+DPM~\cite{fletcher2017power} & Yes \cmark & No \xmark & No \xmark & DVFS + Core hot-plugging \cmark & No \xmark & Voltage level history \xmark & Multicore SoC with Flash \xmark \\
		
		\rowcolor{black!3}AdaMICA~\cite{akhunov2022adamica} & Yes \cmark & Yes \cmark & Yes \cmark & Core activating/deactivating \cmark & No \xmark & Voltage level history \xmark & MSP430FR \xmark \\
		
		\rowcolor{tudCyan!25} 
		\textbf{PEARL} \\ (this work) & \textbf{Yes} \cmark & \textbf{Yes} \cmark & \textbf{Yes} \cmark & \textbf{Core activating/deactivating} \cmark & \textbf{Yes, with NO sleep mode checkpoints} \cmark  & ADC-free, time- and energy-based estimation \cmark & Multicore MCU with SPI FRAM \cmark \\
		\bottomrule
	\end{tabular}
	
	\label{tab:features-comparison}
\end{table*}

\section{Boosting Intermittent Computing}

\label{sec:background} 

Today, many batteryless platforms (e.g.,~\cite{hester2017flicker,desai2022camaroptera,abubakar2022protean,de2020battery}) include single-core MSP430FR series MCUs~\cite{msp430fram}. 
Since FRAM is an embedded component, the processor can access it energy-efficiently by using the optimized interconnecting bus, making these MCUs very efficient for intermittent computing. 
Meanwhile, the MCU market proposes several low-power computing platforms to support artificial intelligence (AI) on resource-limited edge devices. For example, MAX32666~\cite{max32666} has dual Arm Cortex-M4F processors with 1MB Flash and 560KB SRAM. The two ARM Cortex processors can use the optimized interconnecting bus to access the SRAM, Flash, and peripherals. The computational power and parallelism support of MAX32666 come with power efficiency. In active mode, it consumes approximately 3mA at 96MHz. Moreover, in deep sleep mode, it only consumes 10\textmu A. Hence, these MCUs are well-suited for energy-efficient intermittent computing and outperform the MSP430FR series MCUs. 

\subsection{Need for Multicore Intermittent Computing}
\label{sec:use-cases}


Multi-core parallelism enables faster task execution, significantly improving energy efficiency and throughput in batteryless systems~\cite{akhunov2022adamica}. Completing tasks quickly reduces the risk of losing progress during frequent power failures, especially critical when energy bursts are short and workloads are compute-intensive or latency-sensitive. This is particularly valuable in batteryless sensing applications, where timely data processing is essential~\cite{hester2017timely,yildiz2022osdi,yildiz2024adaptable}.

For instance, Camaroptera~\cite{desai2022camaroptera} is a batteryless image sensor that can run on-device AI inference on the images captured by its camera. It can support remote sensing applications, e.g., detecting the presence of the codling moth pest in the fields~\cite{bompani2024accelerating}, by harvesting energy to power a burst of computation: capturing an image and running an object detection model. Cameroptera's current design features a single-core TI MSP430FR series microcontroller. However, due to the compute-intensive nature of on-device AI inference, exploiting multiple cores and distributing the workload across these cores allows the on-device inference to complete faster, reducing mid-task failures and avoiding wasted energy on incomplete inference. In particular, when ambient power is high, a batteryless device can capture more images and perform more inferences via multicore operation, increasing the object detection likelihood and precision. Conversely, when the ambient power is low, accelerating inference computations might compensate for lengthy charging periods, enabling timely detection of the objects and eliminating postponed actions.

\subsection{Prior Art Limitations}

{\bf Design for continuous power.} Existing low-power multicore MCUs, e.g., MAX32666, are not designed for intermittent computing applications, and they do not have embedded FRAM in their architecture. Their main memory is volatile (e.g., SRAM), and they include only Flash as non-volatile memory. Flash memories have high energy requirements, low speed, and limited write endurance (wears out quickly), making intermittent computing on these MCUs unfeasible~\cite{chen2024ifkvs}.


{\bf Heavy backup strategies.} Batteryless systems employ {\em checkpoints }~\cite{choi2022compiler,kortbeek2020time,yildiz2022osdi,ahmed2019efficient} to back up their computational state in non-volatile memory at specific points in time. When the device reboots, it restores its computational state using the last successful checkpoint and resumes the interrupted computation. 
The checkpoint frequency varies depending on the applied checkpoint policy. For instance, QuickRecall~\cite{jayakumar2015quickrecall} uses intermittent voltage checks to decide when to checkpoint. TICS~\cite{kortbeek2020time}  employs a timer-driven checkpointing approach, where a checkpoint is taken at fixed intervals (e.g., 100 ms). 
Differently, RockClimb~\cite{choi2022compiler} is a compiler-driven approach. The compiler splits the program into regions that can fit in the device's capacitor. Before executing a region, the charge in the capacitor is measured. The region is executed only if the stored energy is sufficient to execute it. At the end of each region, a lightweight checkpoint may be taken (e.g., saving only selected registers) based on the register-level dependency between successive regions. Such solutions require frequent non-volatile memory reads and writes, making them extremely inefficient on existing low-power multicore MCUs.   

\subsection{Our Novelties and Differences}

\label{sec:differences-sota}

\sysname introduces power- and energy-aware multicore intermittent computing for the first time, creating the opportunity of making existing ultra-low-power multicore MCUs suitable for intermittent computing. As we introduce in the following sections, adaptive three-threshold voltage tracking brings energy awareness to multicore platforms. Furthermore, \sysname uses an innovative algorithm that exploits voltage tracking and estimates ambient power levels without relying on an analog-to-digital converter (ADC). Using estimated power, \sysname employs a power-aware multicore scaling technique and enables adaptation, increasing the throughput by considering energy-harvesting dynamics.

Table~\ref{tab:features-comparison} compares the main features of \sysname with the relevant prior art. The majority of intermittent computing studies (e.g., ~\cite{yildiz2022osdi,wu2024intos,choi2022compiler}) omit multicore systems. For instance, the DVFS-based (Dynamic Voltage Frequency Scaling) power adaptation technique proposed in~\cite{ahmed2020intermittent,maioli2025dynamic} is device-specific and targets only single-core MCUs for intermittent computing. Other works (e.g., ~\cite{daulby2020improving,akhunov2023enabling}) also employ three-threshold voltage tracking, but they target single-core platforms and propose no dynamic adaptation to sporadic input power and software support to intermittent systems.


There are only a few works on multicore energy-harvesting systems~\cite{balsamo2019momentum,fletcher2017power}. These works propose adaptation to sporadic input power by core-hot-plugging and its combination with DVFS, enabling systems' power-neutrality. 
However, these works use inefficient checkpoints and power-hungry multicore SoCs with Flash. Moreover, they do not address the challenges of multicore intermittent computing, such as task distribution among active cores, state backup and restoration across multiple cores, and task redistribution when the number of active cores changes at runtime. 

The most relevant study, AdaMICA~\cite{akhunov2022adamica},  proposes a multicore intermittent runtime and programming model. It employs power-aware multicore adaptation to reduce the number of power failures and improve the utilization of harvested energy. However, AdaMICA does not support three threshold voltage tracking and pessimistically performs heavy checkpoints (the whole SRAM and registers), making it very inefficient. More importantly, the authors evaluate by simulating/emulating a multicore system equipped with an optimized \emph{internal} FRAM, which does not exist now and limits the applicability of the proposed architecture. AdaMICA strictly requires embedded internal FRAM in multicore architecture, making it incompatible with the off-the-shelf low-power multicore MCUs.

\section{\sysname: System Design}

We design \sysname to enable efficient multicore intermittent computing on ultra-low-power MCUs with small deep sleep mode power consumption. This is made possible through two easy-to-integrate hardware components: an off-the-shelf SPI-based external
fast non-volatile memory module (e.g., ~\cite{spifram}) and a cheap three-threshold voltage tracking circuit~\cite{lukosevicius2017using,akhunov2023enabling}. \sysname software runtime manages these hardware components and ensures responsiveness to dynamic ambient power and harvestable energy to increase throughput. 

\sysname introduces three key concepts that exploit energy and power awareness: (1) avoid unnecessary checkpoints by postponing power failures as long as ambient power allows; (2) avoid using non-volatile memory for computation but only for backups; (3) adapt the system's power consumption to the ambient power strength, using a lightweight monitoring solution. \sysname exploits more parallelism if using many cores improves throughput, considering the available power and backup overheads during intermittent execution.

We consider a dual-core architecture (e.g., MAX32666~\cite{max32666}) in which the main core is responsible for power and energy awareness, activating/deactivating the second core, and checkpoints. The main core uses the external SPI-based FRAM only for storing checkpoints and recovery. As in~\cite{akhunov2022adamica}, the application is composed of parallelizable blocks that can be run in two cores. Programmers annotate these blocks so that \sysname runtime can make decisions to switch between single-core and dual-core modes when executing them.

\subsection{Optimizing Memory Accesses}

\label{sec:energy-levels}

The downside of using an external SPI-based FRAM is that interfacing it through SPI is energy-hungry. Moreover, the FRAM is single-ported, leading to scalability issues for multicore architectures. To overcome this inefficiency, \sysname 
limits FRAM access to only backup and restore operations. Furthermore, \sysname minimizes the frequency of backup and restore operations through
energy awareness by adopting three threshold voltage tracking to multicore MCUs. 

To monitor the energy level in the energy storage capacitor, existing systems (e.g.,~\cite{maeng_supporting_2019,jayakumar2015quickrecall}) either use internal comparators of the MCU or employ a basic monitoring circuitry that keeps track of the voltage level and compares it against two predefined voltage thresholds: one for performing backup and turning off the system ($V_{L}$), and other for turning on the system and performing restoration ($V_{H}$). \sysname incorporates one additional voltage threshold ($V_{M}$) that resides in between the two above~\cite{lukosevicius2017using,akhunov2023enabling}. 

\begin{figure}
    \centering
    \includegraphics[width=1\columnwidth]{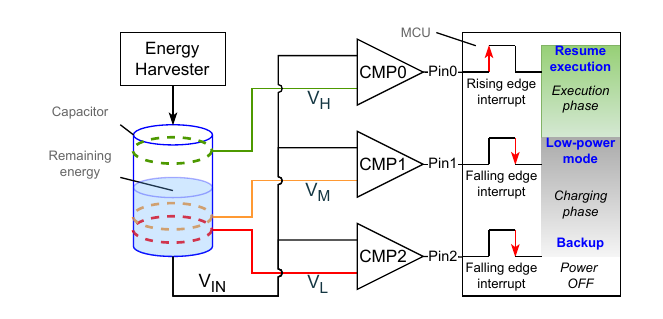}
    \caption{\sysname energy level monitoring strategy.}
    \label{fig:energy-levels}
\end{figure}

Figure~\ref{fig:energy-levels} demonstrates the overview of the energy monitoring. When the capacitor voltage reaches the higher threshold (green line), a comparator\footnote{Evaluates if $V_{in}\ge V_{ref}$, where $V_{in}$ and $V_{ref}$ are the input and reference voltage, respectively.} (CMP0) asserts Pin0, which triggers a rising-edge interrupt in the MCU, denoting that the system has harvested enough energy to restore from the last checkpoint and resume the interrupted computation. Once the capacitor voltage hits the lower threshold (red line), another comparator (CMP2) resets Pin2, triggering the backup procedure and power off. Our system benefits from the third voltage threshold (orange line), triggering a falling-edge interrupt to warn the system to transition to the lowest power mode {\bf retaining} volatile memory content. 

\subsubsection{Low Power Memory Retention Mode (LPMRM).} Staying at this mode, \sysname freezes the ongoing computation but {\bf keeps the volatile computational state, eliminating the need for checkpoints}, and waits for the capacitor to reach either $V_{H}$ or $V_{L}$, consuming power in the order of micro-watts or nano-watts depending on the device. If ambient power ($P_{amb}$) exceeds the LPMRM power consumption ($P_{LPM}$), the energy in the capacitor reaches a higher threshold in time:
\begin{align} 
  T_{H}=\frac{E_{H-M}}{P_{amb} - P_{LPM}},
\end{align}
where $E_{H-M}$ is the energy stored between $V_{H}$ and $V_{M}$. Conversely, if $P_{amb}<P_{LPM}$ holds, time to reach $V_{L}$ is:
\begin{align} 
  T_{L}=\frac{E_{M-L}}{P_{LPM} - P_{amb}},
\end{align}
where $E_{M-L}$ is the energy stored between $V_{M}$ and $V_{L}$. Both equations show that as long as $P_{amb}=P_{LPM}$, \sysname can stay at LPMRM, suspending expensive checkpoints and eliminating external non-volatile memory access. Consequently, \sysname backups only when the stored energy in the capacitor drops below the backup
threshold and the capacitor continues to discharge since the
ambient power is smaller than the LPMRM  power consumption of the multicore platform. 

\subsubsection{Defining Thresholds.} $V_{H}$ is fixed and defined at the system design stage, and it also depends on the capacity of the energy buffer and the voltage requirements of the load. $V_{L}$ is adjustable by software and corresponds to the energy necessary to perform a checkpoint, which has a different cost for single-core and dual-core modes. $V_{M}$ is also controlled by software and reflects $E_{M-L}$ needed to transition to LPMRM and arbitrary time to wait for backup. Note that to guarantee the execution progress, $E_{H-M}$ must be enough to restore after power is on and to execute a part of the given application. 



\subsubsection{SRAM Retention.} \sysname uses SRAM as the main memory, and does not permit programs to use the external FRAM. More precisely, \sysname forces the off-the-shelf multicore systems to exploit the energy efficiency of their de facto architecture. This strategy {\bf eliminates external FRAM access}, increasing the efficiency of, in particular, memory-bound operations significantly. \sysname uses FRAM exclusively for backing up the computational state. Checkpoints are taken only when the remaining voltage drops below $V_{L}$. In this case, when a power failure is unavoidable, \sysname moves interim computation results and sensitive architectural state from SRAM to non-volatile memory. While charging in LPMRM, \sysname retains volatile memory contents and avoids expensive checkpoints. This strategy increases performance, reduces execution energy consumption, and eliminates the need for compiler analysis (e.g.,~\cite{yildiz2022osdi,choi2022compiler}) or task-based program transformation (e.g.,~\cite{maeng2017alpaca}).

\subsubsection{Checkpoints.} Our checkpoints back up the entire SRAM and processor registers. 
As a basic optimization, we split SRAM into three equal sections: one section shared among two cores and two individual sections for ongoing computations on cores. This splitting allows for a constant cost for checkpoints for both single-core and dual-core modes. When the system is operating in single-core mode, only the individual section of the main core and the shared section are checkpointed. As shown in Section~\ref{sec:results}, \sysname significantly outperforms existing approaches even with our current checkpointing strategy.

\subsection{Power-Aware Architectural Scaling}
\label{sec:power-pheromone}

Being aware of ambient power strength is essential for \sysname to adapt the multicore architecture accordingly. 
When \sysname encounters a parallelizable block in the application, it considers the available power and makes a decision on the multicore mode, presented briefly as follows:
\begin{align} 
      Mode = 
    \begin{cases}
        2C, & \text{if } P_{amb}\ge P_{exec2C} \\
        1C, & \text{otherwise},
    \end{cases}
    \label{eq:modes}
\end{align}
where $P_{amb}$ is the ambient power value, $P_{exec2C}$ is the power consumption of the dual-core execution mode, and 2C and 1C are dual-core and single-core modes, respectively.

One trivial solution for tracking the ambient power is sampling input power levels on demand. However, this solution leads to an instantaneous switch between modes on very short power spikes and drops. AdaMICA~\cite{akhunov2022adamica} uses a smoother switching technique relying on the input power history. First, it collects the sampled values in a finite buffer; then, at the beginning of a parallelizable block, it calculates the mean of these values and compares it with $P_{exec2C}$, which can be obtained from datasheets or energy profiling. 
However, this solution requires intensive ADC sampling, memory space for holding the historical data, and mean computation. 

\subsubsection{Algorithm Overview.} \sysname uses a computationally lightweight and energy-efficient approach that does not require ADC sampling and large memory space for making adaptive decisions. The monitoring circuitry (Figure~\ref{fig:energy-levels}) triggers interrupts upon energy level changes. \sysname keeps track of the {\em time between these interrupts} to obtain ambient power estimates, which introduces significantly smaller energy costs compared to individual ADC measurements. To capture the {\em ambient-power trend} and keep the system in an optimal multicore configuration state, \sysname employs Exponentially Weighted Moving Average (EWMA) on the obtained power estimates. EWMA offers a computationally lightweight solution with a very small memory footprint and sufficient performance. By using the predicted power level, \sysname calculates the estimated throughput for different multicore execution modes in a computationally lightweight manner and decides on the most performant configuration. 

\begin{figure*}
    \centering
    \includegraphics[width=0.9\textwidth]{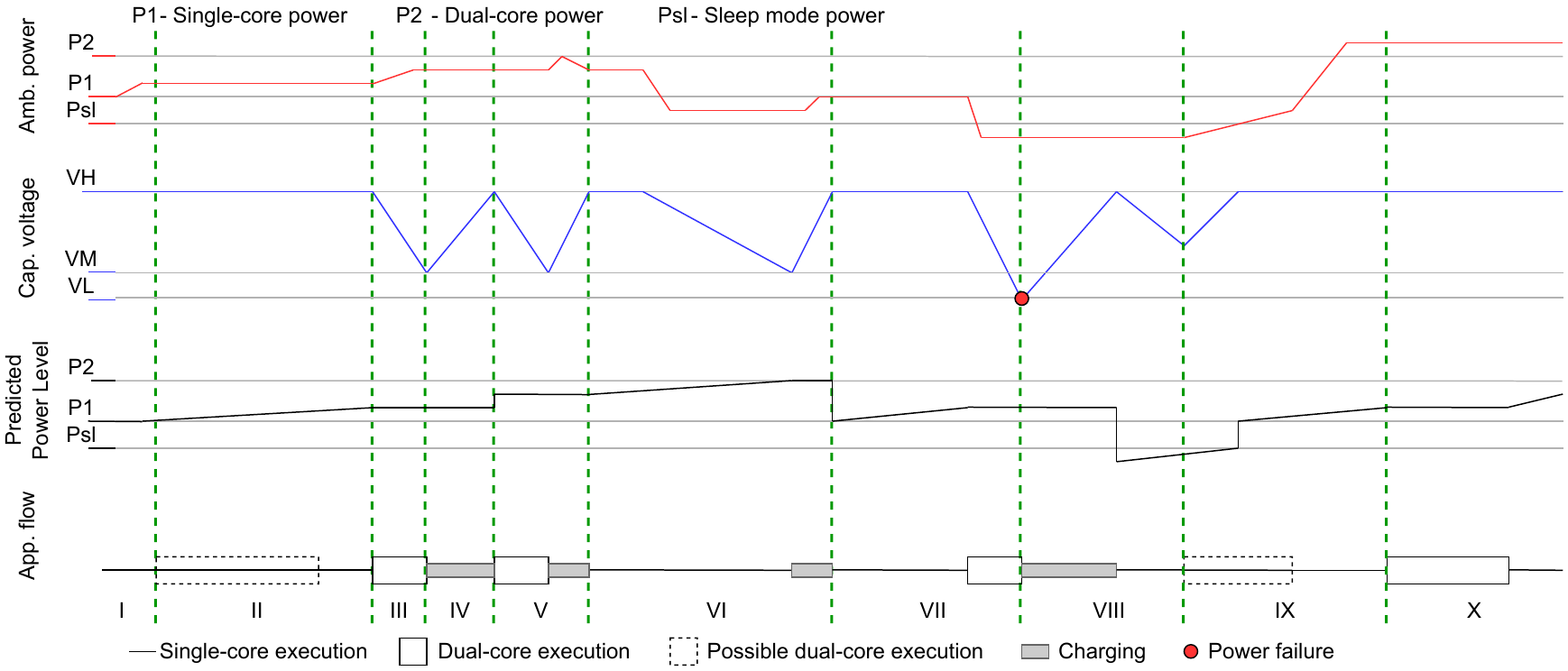}
    \caption{The example of execution flow and power level prediction.}
    \label{fig:pearl-example}
\end{figure*}

\subsubsection{Algorithm Description.} 
The algorithm requires the power consumption of single-core (1C) and dual-core (2C) modes, denoted by $P_{1C}$ and $P_{2C}$, which can be obtained either from datasheets or experimental measurements. 

\noindpar{Obtaining Power Estimates Almost for Free.} Assume that \sysname starts execution at time $t_0$ with a fully charged capacitor. \sysname continues execution by switching between different modes (depending on whether the code is parallelizable or not) until it is interrupted when the voltage hits the medium threshold $V_M$ at time $t_1$. Let $\Delta t_{1C}$ and $\Delta t_{2C}$ denote the time spent in 1C and 2C modes in this time interval, i.e., in $[t_0, t_1]$. Note that the capacitor is also being charged by the ambient power denoted by $P_{amb}$ during this period. {\color{black}  However, the capacitor {\em discharges} since $P_{1C}>P_{amb}$ holds most of the time, and the voltage value of the capacitor drops from $V_H$ to $V_M$. {\bf Without any ADC measurement}, at time $t_1$, \sysname can obtain an estimate of the average ambient power in the time interval $[t_0, t_1]$  as follows:
\begin{equation}
    \hat{P}_{amb}^{(1)} = \frac{\Delta t_{1C}P_{1C}+\Delta t_{2C}P_{2C}-E_{H-M}}{\Delta t_{1C} + \Delta t_{2C}}
\end{equation} 
where  $E_{H-M}$ is the capacitor energy, as defined in Section~\ref{sec:energy-levels}, which is known in advance; and the MCU obtains $\Delta t_{1C}$ and $\Delta t_{2C}$ by reading its internal timers. It is worth mentioning that the MCU switches to sleep mode at time $t_1$ after obtaining $\hat{P}_{amb}^{(1)}$ to charge its capacitor up to $V_H$.
}

Assume that the ambient power is greater than the sleep mode power consumption of the device, i.e., $P_{LPM}<P_{amb}$. Then, after hitting $V_M$ at time $t_1$, the capacitor will be {\color{black}\em charged} (from voltage level $V_M$) until its voltage level hits $V_H$ at time $t_2$. {\color{black} MCU can record  $t_1$ and $t_2$ since it receives an interrupt from the energy monitoring circuitry presented in Figure~\ref{fig:energy-levels} at these times and can read its internal timer. After waking up from deep sleep mode at the time $t_2$, {\bf without any ADC measurement}, \sysname can obtain the estimate for the average ambient power in the time interval  $[t_1, t_2]$  as:
\begin{equation}
    \hat{P}_{amb}^{(2)} = \frac{E_{H-M}}{t_2-t_1}.
\end{equation} 
}

{\color{black}\noindpar{Predicting Ambient Power Level.}
At the time $t_2$, using the power estimates $\hat{P}_{amb}^{(1)}$ and $\hat{P}_{amb}^{(2)}$, \sysname estimates the historical summary of the  power by using EWMA as follows:
\begin{equation}
    \hat{\bf P} = (1-\alpha)\hat{\bf P} +  \alpha(\hat{P}_{amb}^{(1)} +  \hat{P}_{amb}^{(2)})/2,
    \label{eq:pheromone-update}
\end{equation}
where $\alpha$ is the weight for the update. While computationally lightweight, EWMA is robust and adaptive to short-term variations~\cite{kansal2007power,geissdoerfer2019getting}. \sysname uses $\hat{\bf P} $ as an indicator of {\em future} ambient power level. When the ambient energy is high, the voltage level might not hit $V_M$ for a long time. In this case, \sysname would not get any power estimates and update $\hat{\bf P}$. To prevent this situation, \sysname uses a periodic timer to gradually increment the predicted power level $\hat{\bf P}$. {\color{black} The value of the timer depends on the execution mode and is calculated at the design stage. For example, the timer for $\hat{\bf P}$ updates in single-core mode is set as $T_{1C}=E_{H-M}/P_{1C}$. That is, if the single-core execution did not reach $V_{M}$ in time $T_{1C}$, it means that $P_{amb}>P_{1C}$ and $\hat{\bf P}$ can be updated. Given the dynamism of ambient power and the benefits that the architectural scaling brings (see Section~\ref{sec:results}), the overhead of these rare timer interrupts is negligible.}  
}

{\color{black}
\noindpar{Lightweight Decision Making.} 
\sysname uses {\em throughput} as a simple heuristic to decide which configuration to switch to by considering the predicted power level. It considers three cases to decide on which configuration to select.

{\par \bf \em Case-0.} If the predicted power level, represented by $\hat{\bf P} $, is greater than the power requirements of dual-core execution mode, i.e.,  $\hat{\bf P} > P_{2C}> P_{1C}$, this means that the power is strong with no or rare power failures and it is safe to switch to a more performing mode (i.e., from 1C to 2C).
 
{\par \bf \em Case-1.} Assume that $\hat{\bf P} < P_{1C}<P_{2C}$ holds, which means that the predicted power level $\hat{\bf P}$ is smaller than both the power consumption of dual-core and single-core execution modes. In this case, until the voltage level hits $V_M$ threshold, \sysname can spend either $\Delta t_{1C}$ time in 1C mode or $\Delta t_{2C}$ time in 2C mode, presented as follows:
\begin{equation}
    \Delta t_{1C} = \frac{E_{H-M}}{P_{1C}-\hat{\bf P}}, \quad \Delta t_{2C} = \frac{E_{H-M}}{P_{2C}-\hat{\bf P}}.
\end{equation}
Since 2C mode can perform 2 times more operations compared to 1C mode, the ratio of the computational throughput in 2C mode (i.e, $Th_{2C}$) with respect to the 1C mode (i.e, $Th_{1C}$) can be represented as: 
\begin{equation}
    {\bf Th} = \frac{Th_{2C}}{Th_{1C}} = \frac{2 \times \Delta t_{2C}}{\Delta t_{1C}} = \frac{2(P_{1C}-\hat{\bf P})}{P_{2C}-\hat{\bf P}}.\label{eq:throughput}
\end{equation} In this case, if ${\bf Th} \geq 1$ then 2C is more performant and \sysname switches to 2C mode; otherwise it keeps 1C execution mode.   
{\par \bf \em Case-2.} Assume that $ P_{1C}< \hat{\bf P} < P_{2C}$ holds. In this case, the system will never hit $V_M$ during 1C execution mode since  $ P_{1C}< \hat{\bf P}$ holds. On the other hand, in 2C mode, the system will compute until it hits the $V_M$ threshold, which will lead to a charging time until the system hits $V_H$ again. To compare the throughput in 2C mode with respect to that in 1C mode, we need to take into account also the charging time from $V_M$ to $V_H$ when \sysname operates in 2C mode. Therefore: 
\begin{equation}
   \Delta t_{2C} = \frac{E_{H-M}}{P_{2C}-\hat{\bf P}} + \frac{E_{H-M}}{\hat{\bf P}}, 
\end{equation}
where the right component captures the charging time in 2C mode. Now, the throughput ratio can be calculated in a slightly different way. The reason is that in 1C mode, we will have continuous computation during the whole $\Delta t_{2C}$. On the other hand, in 2C mode, the system can perform computation only during active mode, excluding the charging time (i.e., only during $\frac{E_{H-M}}{P_{2C}-\hat{\bf P}}$). Therefore, 
\begin{equation}
    Th_{1C} = \frac{E_{H-M}}{P_{2C}-\hat{\bf P}} + \frac{E_{H-M}}{\hat{\bf P}}, \quad Th_{2C} = 2 \times \frac{E_{H-M}}{P_{2C}-\hat{\bf P}}
    \label{eq:case2-threshold}
\end{equation}
Putting these values in Equation~\ref{eq:throughput}, \sysname can decide to switch to 2C if ${\bf Th} \geq 1$, or vice versa.
}

\subsubsection{How it Works.} 

In Figure~\ref{fig:pearl-example}, we show the dynamics of the ambient power, capacitor voltage, and the predicted power level during the execution of a representative parallelizable computational workload. We split the execution into ten parts to analyze different scenarios of maintaining the power awareness of PEARL. 

(I) Initially, the execution mode is single-core, the predicted power level is equal to $P_{1C}$, and the capacitor voltage is at the highest level. The predicted power level is gradually increased during single-core execution, while the capacitor voltage remains at the same level because ambient power is equal to or greater than the power consumption of the single-core mode. (II) Then, the application encounters a parallelizable code block (dashed rectangle), but the execution is not switched to the dual-core mode because the predicted power level is not sufficient. (III) At a certain point, the program encounters another parallel block and executes it in the dual-core mode. The capacitor starts being depleted because ambient power is lower than the dual-core power consumption. However, the predicted power level remains intact because it is updated according to Equation~\eqref{eq:pheromone-update} only at each wakeup. (IV) When the capacitor voltage reaches VM, the execution is interrupted to transition to sleep mode and charge the capacitor. (V) Once the device wakes up, the predicted power level is updated. (VI) When execution returns to the single-core mode, the level starts gradually increasing.
(VII, VIII) When ambient power becomes lower than the power consumption of sleep mode, a power failure is inevitable. After the power failure and charging, the predicted power level is updated and becomes correspondent with the actual ambient power level. (IX) This power level
cannot maintain the dual-core mode; therefore, the program continues with the single-core mode. (X) When a parallelizable block is encountered again, the predicted power reaches a sufficient level to switch to benefit from the dual-core mode. This example shows that even if there exists a certain divergence between the predicted power and actual ambient power levels, the proposed power-awareness technique demonstrates reasonable adaptation accuracy.

\subsubsection{Scalability.} \sysname's power-aware scaling strategy can be easily extended to many cores by incorporating the power consumption of a desired multicore mode into Equations~\eqref{eq:modes} -~\eqref{eq:case2-threshold}. For instance, in a 4-core system, $\hat{P}_{amb}^{(1)}$ can be estimated by adding the power consumption and time of all the modes involved in a particular execution. The throughputs (Th) for switching to additional cores can also be easily calculated using equations for single-core and dual-core configurations. It is worth mentioning that compared to AdaMICA~\cite{akhunov2022adamica}, \sysname's architectural scaling is lightweight in terms of sampling power, computation, and memory requirements, with a good performance as we show in Section~\ref{sec:results}.

\section{\sysname: Implementation}
\label{sec:implementation}

This section describes the implementation details of the \sysname prototype. As hardware, \sysname requires a dual-core low-power MAX32666 platform, a cheap voltage monitoring circuitry, an external SPI FRAM, and an additional RF energy-harvesting kit. 

\begin{figure}
    \centering
    \includegraphics[width=0.8\columnwidth]{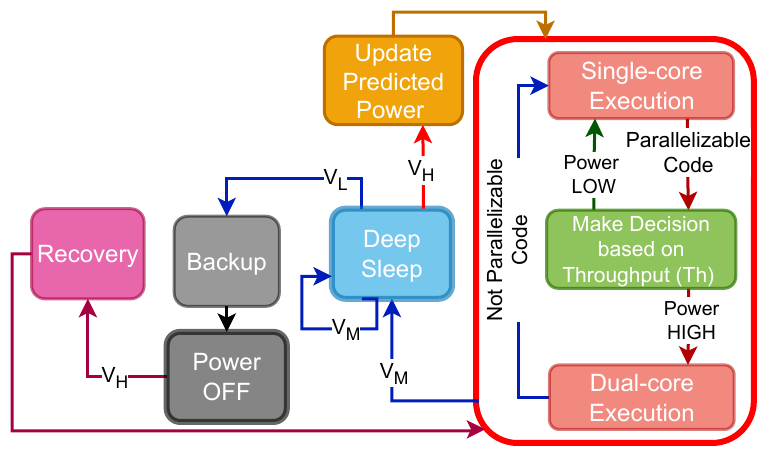}
    \caption{FSM of the \sysname  computational flow.}
    \label{fig:pearl-fsm}
\end{figure}

\subsection{\sysname Intermittent Computing Flow}

\sysname follows the computing flow shown in Figure~\ref{fig:pearl-fsm}. 
The application starts in single-core mode when the capacitor is fully charged. When the program encounters a parallelizable code, \sysname checks the predicted power level. If it is low, the execution continues in single-core mode; otherwise, \sysname splits the computational load between two cores and configures the voltage monitor to adjust the medium ($V_M$) and lower ($V_L$) voltage thresholds accordingly. It is worth mentioning that $V_M$ and $V_L$ in single-core mode are different than those in dual-core mode due to the different power consumption and energy requirements for checkpoints. 

The execution can switch back to the single-core mode when a non-parallelizable code is encountered or when the predicted power is low. In these cases, \sysname transitions to deep sleep mode if $V_M$ is triggered and waits to charge the capacitor. If the input power is higher than deep sleep mode power consumption, the capacitor is charged to $V_H$, the predicted power level is updated (see Section~\ref{sec:power-pheromone}), and the computation resumes either in the single- or dual-core mode. Otherwise, the voltage in the capacitor reaches $V_L$, and \sysname backs up SRAM and turns off the power. The interrupted computation on MAX32666 is restarted when the capacitor hits $V_H$. The execution can finish in any mode.

\subsection{\sysname Voltage Tracking}
\label{sec:hml-impl}

We use the three-threshold voltage tracking circuit in (see Section~\ref{sec:energy-levels}), which includes a storage capacitor to power the multicore MCU, a voltage tracker circuit composed of nano-power comparators, and a voltage divider built using an ultra-low-power programmable digital potentiometer. We set $V_{L}$, $V_{M}$, and $V_{H}$ as 2V, 2.5V, and 2.9V, respectively. Thus, \sysname software can configure the $V_{L}$ and $V_{M}$ from 2V to 2.5V and from 2.5V to 2.9V, respectively, by changing the potentiometer divider rates via I2C. The reconfigurability of $V_{L}$ and $V_{M}$ is crucial since these thresholds are different in single-core and dual-core operation modes. The circuit power consumption is approximately 7.8\textmu W, which is two orders of magnitude lower than the ADC power consumption of MAX32666, which is approximately 690\textmu W. Note that the total energy consumption of the ADC depends on the sampling frequency, as we evaluate in Section~\ref{sec:hardware-results}. 

\begin{figure}
   \centering
   \includegraphics[width=0.65\columnwidth]{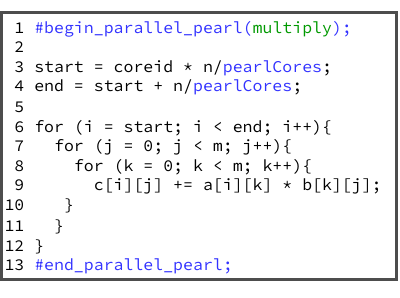}
   \caption{\sysname software library code example.}
   \label{fig:code-example}
\end{figure}

\subsection{\sysname Software Library}
\label{sec:sw-lib}

Figure~\ref{fig:code-example} shows a parallelizable code annotation using the \sysname library. We modified the lightweight C library presented in~\cite{akhunov2022adamica} to ease the integration of \sysname to applications. Developers annotate parallelizable sections of code using \texttt{begin\_parallel\_pearl} and \texttt{end\_parallel\_pearl} macros. These macros define a C function for the parallel task, reconfigure the multicore system at runtime, and manage communication between the main and secondary cores. Secondary cores notify the main core upon task completion, and the main core proceeds only after all parallel tasks have finished.

{\bf  Interrupt Handlers.} First, we added an interrupt handler for the medium voltage threshold. When the interrupt is triggered, MAX32666 transitions to deep sleep mode, retaining the content of volatile memory (SRAM and registers). We set two interrupt triggers as events for waking up from the deep sleep mode: reaching the $V_{H}$ threshold to resume interrupted computation and reaching the $V_{L}$ threshold to back up and power off the board. 

{\bf Power Prediction.} Second, we dedicated one 8-bit variable to store the current predicted power level. To read and modify this variable, we added getter and setter functions.

{\bf Multicore Scaling.} Third, we simplified the decision-making algorithm for switching between architectural configurations by employing only a call to the getter function and one comparison operation for the predicted power level (see our algorithm in \Cref{sec:power-pheromone}). With this modification, there is no longer a need for an AdaMICA-like decision-making table and power history, which frees up space in memory. It is worth mentioning that $V_{L}$ and $V_{M}$ thresholds are also adjusted by configuring the programmable potentiometer on the voltage tracking circuitry.

{\bf Multicore Backup.} Finally, we fixed different SRAM regions to back up single-core and dual-core modes. We specified two base address pointers in C to the beginning of SRAM (\texttt{0x20000000}) for the dual-core mode and to the $\frac{1}{3}$ of SRAM (\texttt{0x2005D554}) for the single-core mode. Our library uses these base address pointers and two dynamic pointers to restrict the usage of SRAM by different modes to the dedicated memory regions. Therefore, the programmer must use the provided attributes for declaring variables, \texttt{\_pearl\_1c} or \texttt{\_pearl\_2c}, which map variables to the corresponding memory regions. Upon a checkpoint instruction, \sysname checks the currently running computational mode and selects the starting address of SRAM to back up accordingly. 

\section{Evaluation}
\label{sec:results}

For comparison, we consider two state-of-the-art approaches, AdaMICA~\cite{akhunov2022adamica} and RockClimb~\cite{choi2022compiler}. To make a fair comparison, we use SRAM as the main memory and FRAM as the memory only for backups for all solutions. We perform both simulations and testbed experiments.

\subsection{Simulation Setup} We run an in-house simulation in Python to estimate execution time and energy consumption, using parameters (see Table~\ref{tab:sys-params}) obtained from the MAX32666~\cite{max32666} datasheet. The application comprises a set of randomly distributed instructions with different time and energy costs, {\color{black}which are estimated based on the MCU frequency (96MHz) and active power (10mW), respectively}. The instructions are grouped into multiple non-parallelizable and parallelizable blocks randomly shuffled. We {\color{black}assume 100\% and} 80\% of parallelizable blocks in our application. To execute the application, we go through the instructions and drain the capacitor according to the instruction energy cost. We run the application under a variety of ambient power levels, which dictate the intensity of charging. We periodically check the energy level in the capacitor to take action corresponding to a particular voltage threshold. Furthermore, we periodically increase the level of predicted power during computation and update the level when the system wakes up. The predicted power level is then used to decide on parallelizable code blocks. The parallelizable blocks are executed in many iterations, at the beginning of which the decision is made. 

\begin{table}
  \centering
  \scriptsize
    \caption{List of the simulation parameters.}
  \begin{tabular}{m{3cm}m{1.0cm}m{2.6cm}m{1.1cm}}
    \toprule
    \rowcolor{black!3}
    \textbf{Parameter} & \textbf{Value} & {\bf Parameter} & \textbf{Value}\\
    \midrule
    Instruction count & 1 000 000 & Voltage compar. time & 0.0624 \textmu s\\
    \rowcolor{black!3}
    Capacitor & 1 mF & Voltage compar. energy & 0.68 nJ\\
    Active power & 10 mW & Checkpoint time & 96350 \textmu s \\
    \rowcolor{black!3}
    Deep sleep power (LPMRM) & 0.033 mW & Checkpoint energy & 829610 nJ\\
    \rowcolor{black!3}
    Boot time & 500 \textmu s & Recovery time & 96350 \textmu s\\
    Boot energy & 5450 nJ & Recovery energy & 829610 nJ\\
    \rowcolor{black!3}
    Core-to-Core time & 0.052 \textmu s & FRAM write energy & 829610 nJ\\
    Core-to-Core energy & 0.576 nJ & FRAM read energy & 829610 nJ\\
    
    \bottomrule
  \end{tabular}
  \label{tab:sys-params}
\end{table}

\begin{figure}
    \centering
    \includegraphics[width=0.8\columnwidth]{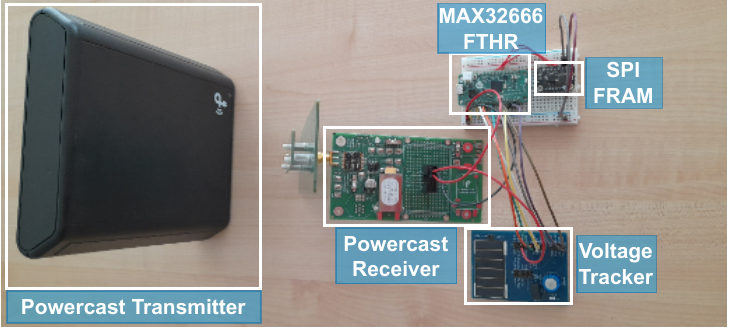}
    \caption{Real hardware evaluation setup.}
    \label{fig:hw-setup}
\end{figure}

\subsection{Testbed Setup} To validate the simulation results, we deploy a real-hardware evaluation setup based on the MAX32666 platform, as shown in Figure~\ref{fig:hw-setup}. 
The platform features 560KB of SRAM shared between two Cortex-M4 cores running at 96MHz. To support intermittent computing, we augment the platform with an external 512KB SPI-based FRAM used for backups. 

As an energy-harvesting part, we employ the Powercast TX91501-3W to generate RF waves and the P2110-EVB to collect RF energy in an internal 50mF capacitor and convey the harvested energy to MAX32666. The board is connected via I2C to the voltage monitoring circuitry to receive three signals for all the thresholds and to send signals adjusting $V_{M}$ and $V_{L}$ thresholds to a desired mode. The energy-harvesting kit is equipped with a booster converter that keeps the supplied voltage from the capacitor to the MCU stable at 3.3V. However, the actual voltage output from the capacitor varies from 2V to 1.02V. We set 1.9V as the $V_{H}$ threshold, 1.35V and 1.25V as the $V_{M}$ and $V_{L}$ thresholds for the single-core mode, and 1.45V and 1.35V as the $V_{M}$ and $V_{L}$ thresholds for the dual-core mode, respectively. We empirically verified these values, ensuring that $V_{L}$ and $V_{H}$ guarantee checkpoint and recovery procedures with no power interrupts, and that there is enough energy between $V_{M}$ and $V_{L}$ to handle the interrupt and switch the system to sleep mode. We set the timer for $\hat{P}$ updates at single-core mode ($T_{1C}$) equal to 4.5s. Note that both SPI and I2C are standard communication protocols for MCUs and require almost no effort to set up. 

\subsubsection{Computational Workload.} 
We repetitively execute a convolution operation on a 32$\times$32 single-channel image, applying a 2$\times$2 kernel. 
We count the number of multiply-and-accumulate (MACs) performed in 60 seconds, calculate the number of MAC operations per second (MACOPS), and use this measure as a performance metric. The metric is indicative because it takes into account time spent on actual computation, on charging, and on waiting for available power (i.e., on the power-off state). 

\subsubsection{Voltage tracking.}
We compare the voltage tracking circuit with the ADC-based voltage tracking approach regarding the response time to voltage drops and energy consumption. Since the performance of the ADC-based approach depends on the sampling rate, we consider four sampling rates in ksps: 0.1, 1, 4, and 7.8. We utilize a digital oscilloscope to monitor voltage drops across the capacitor and a logic analyzer to track the trigger signal generated by the voltage tracking circuit and ADC interrupt.

\subsubsection{Real Application Scenario.} We use the testbed to execute the \emph{plant monitoring} application presented in ~\cite{akhunov2022adamica}, which uses two different cameras, RGB and thermal, to detect plant disease via CNN classification. By default, the RGB camera is powered by a solar panel. The thermal camera turns on only when an additional RF power source is applied. Hence, more samples can be processed when ambient power is higher. 

\subsection{Simulation Results} 

\begin{figure}
    \centering
    \begin{subfigure}[b]{\columnwidth}
        \centering\includegraphics[width=1\columnwidth]{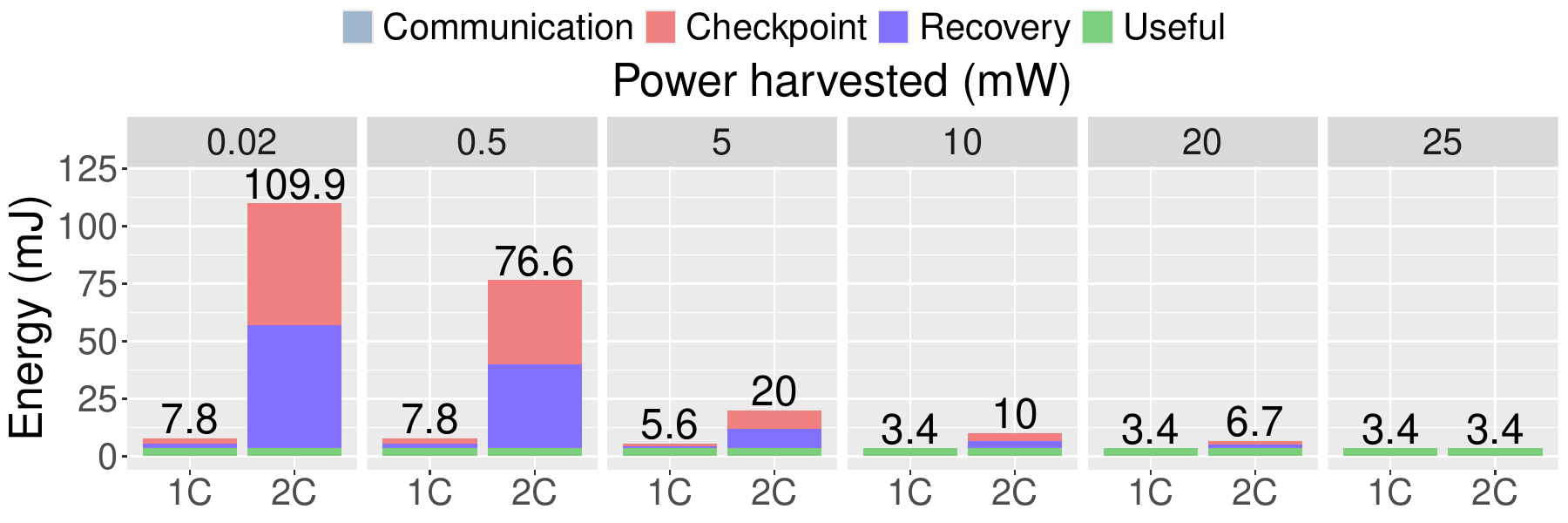}
        \caption{AdaMICA}
        \label{fig:energy-jit}
    \end{subfigure}
    \begin{subfigure}[b]{\columnwidth}
        \centering\includegraphics[width=1\columnwidth]{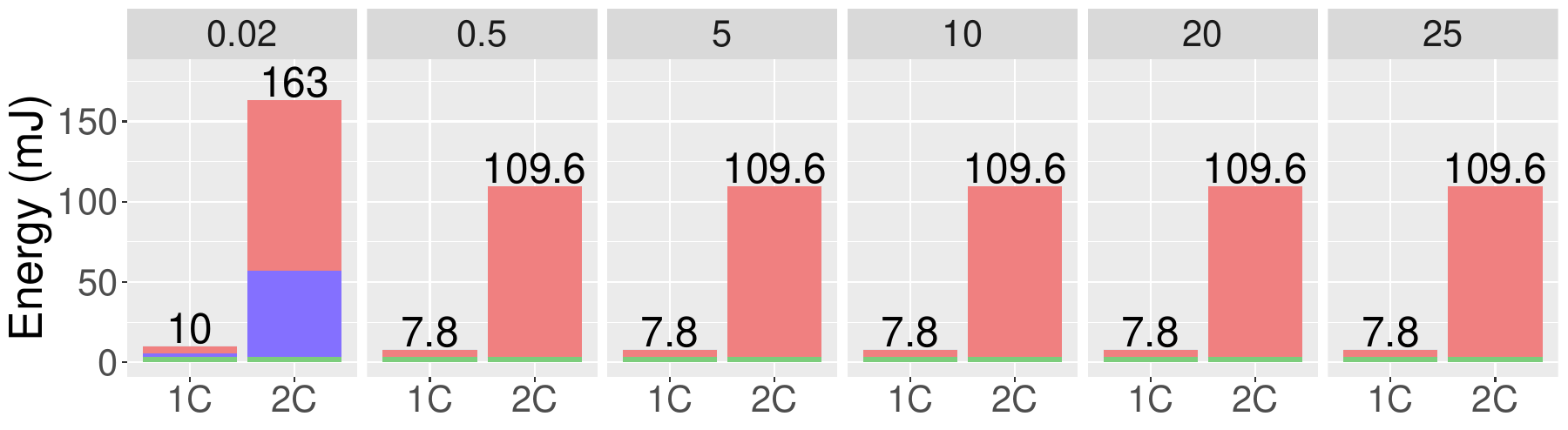}
        \caption{RockClimb}
        \label{fig:energy-rockcl}
    \end{subfigure}
    \begin{subfigure}[b]{\columnwidth}
        \centering\includegraphics[width=1\columnwidth]{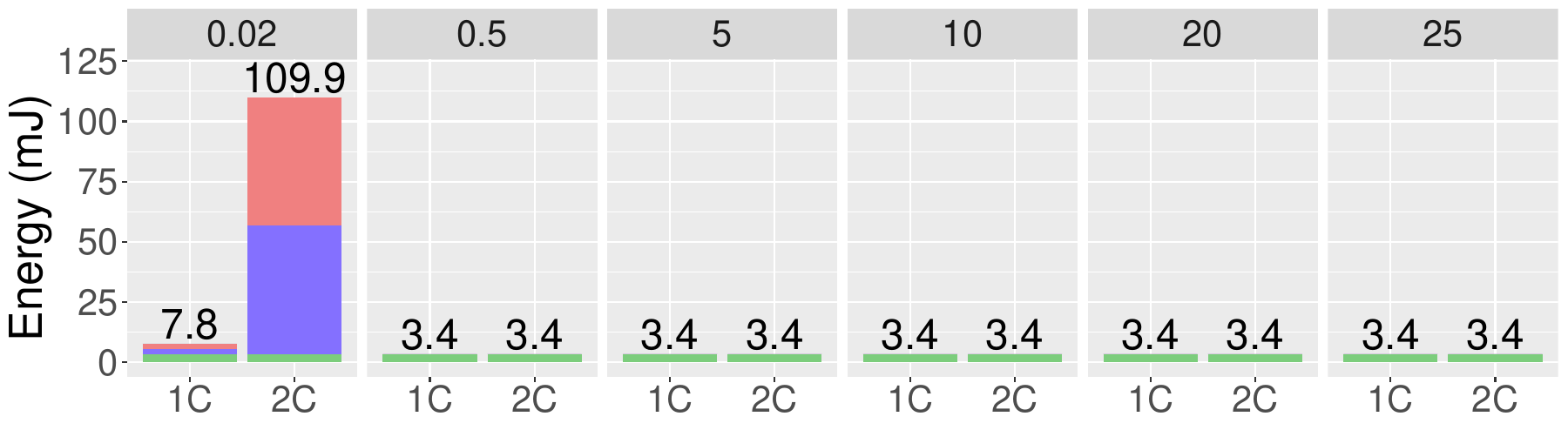}
        \caption{\sysname}
        \label{fig:energy-jit-optim}
    \end{subfigure}
    \caption{Energy consumption comparison.}
    \label{fig:energy-evaluation}
\end{figure}

First, we simulate the code with 100\% parallelization, where the entire application can be equally distributed among two cores. In Figure~\ref{fig:energy-evaluation}, we show the breakdown of energy consumption of AdaMICA, RockClimb, and \sysname. We test environments with different constant levels of ambient power, from 0.02 to 25mW. As seen, when power is lower than the LPMRM power consumption (i.e., 0.02 $<$ 0.033, see Table~\ref{tab:sys-params}), AdaMICA and \sysname exhibit identical energy characteristics due to exploiting a just-in-time checkpoint strategy. However, RockClimb consumes 28\% and 48\% more energy for 1C and 2C modes, respectively. Increasing ambient power to more than the LPMRM power consumption leads to a linear decrease in AdaMICA energy consumption. However, RockClimb benefits only from eliminating recovery overhead because the checkpoints are predefined and performed before the system decides whether to power off or switch to power down mode and wait for charging~\cite[see Section V]{choi2022compiler}. This approach does not allow the system to avoid checkpoints and keeps the energy consumption at the same level even for higher ambient power. Conversely, \sysname spends energy only on useful computation, avoiding all the checkpoint and recovery overheads thanks to the three-threshold voltage tracking approach, which never reaches the lowest voltage threshold with higher ambient power. Overall, \sysname improves energy efficiency by up to 22$\times$ and 32$\times$ compared to AdaMICA and RockClimb, respectively, across the entire power range.

\begin{figure}
    \centering
    \begin{subfigure}[b]{\columnwidth}
        \centering\includegraphics[width=1\columnwidth]{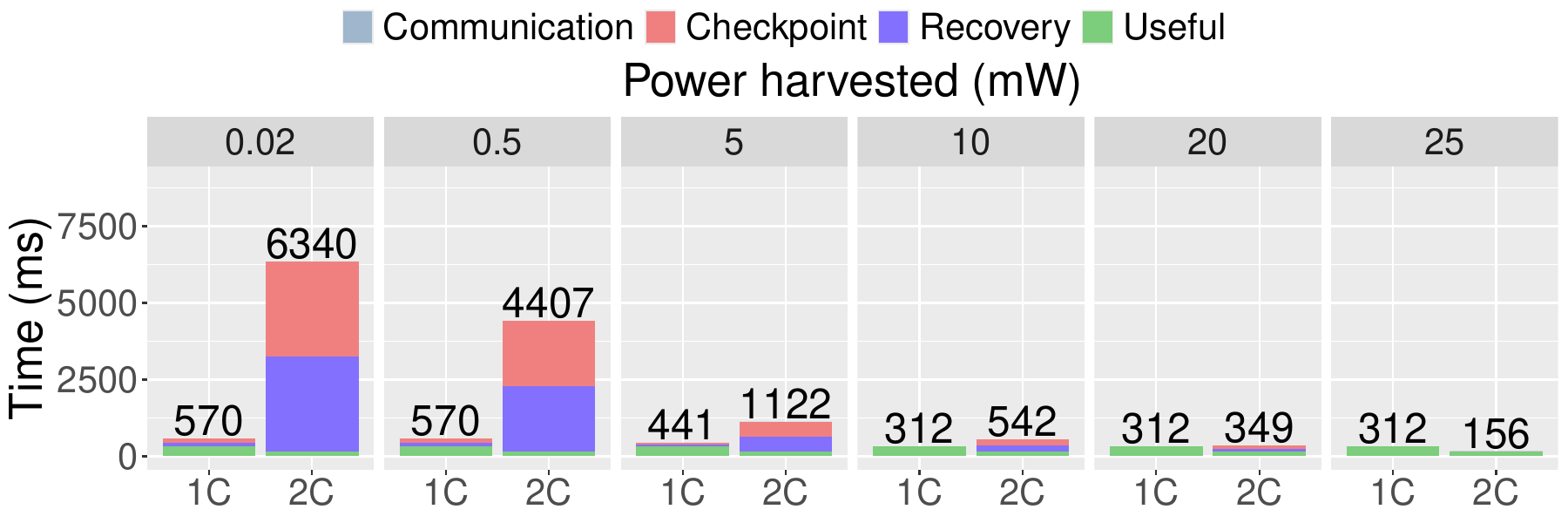}
        \caption{AdaMICA}
        \label{fig:time-jit}
    \end{subfigure}
    \begin{subfigure}[b]{\columnwidth}
        \centering\includegraphics[width=1\columnwidth]{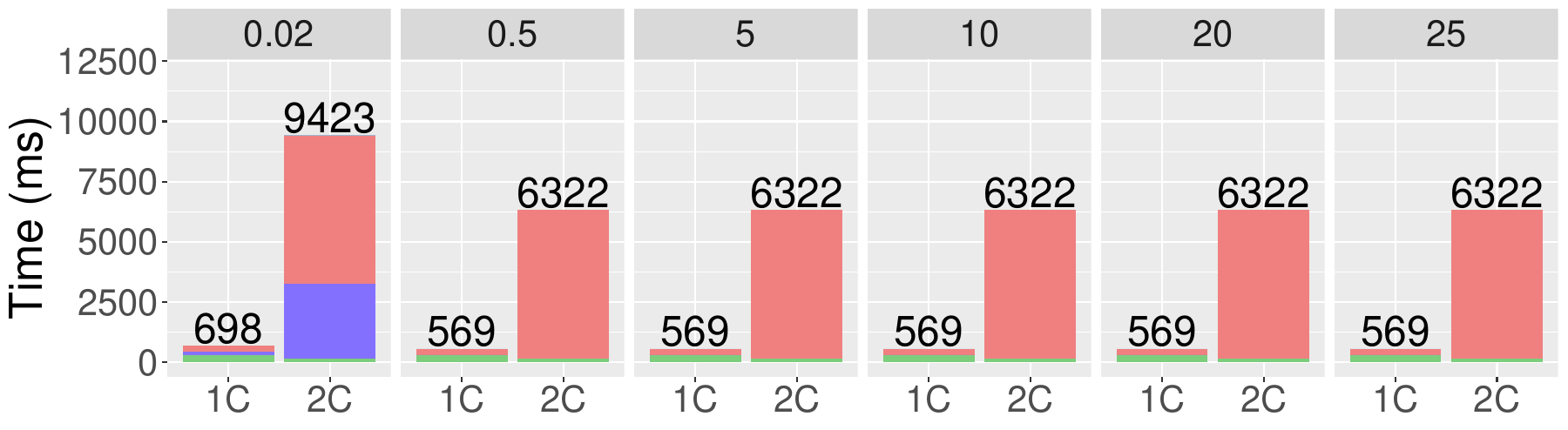}
        \caption{RockClimb}
        \label{fig:time-rockcl}
    \end{subfigure}
    \begin{subfigure}[b]{\columnwidth}
        \centering\includegraphics[width=1\columnwidth]{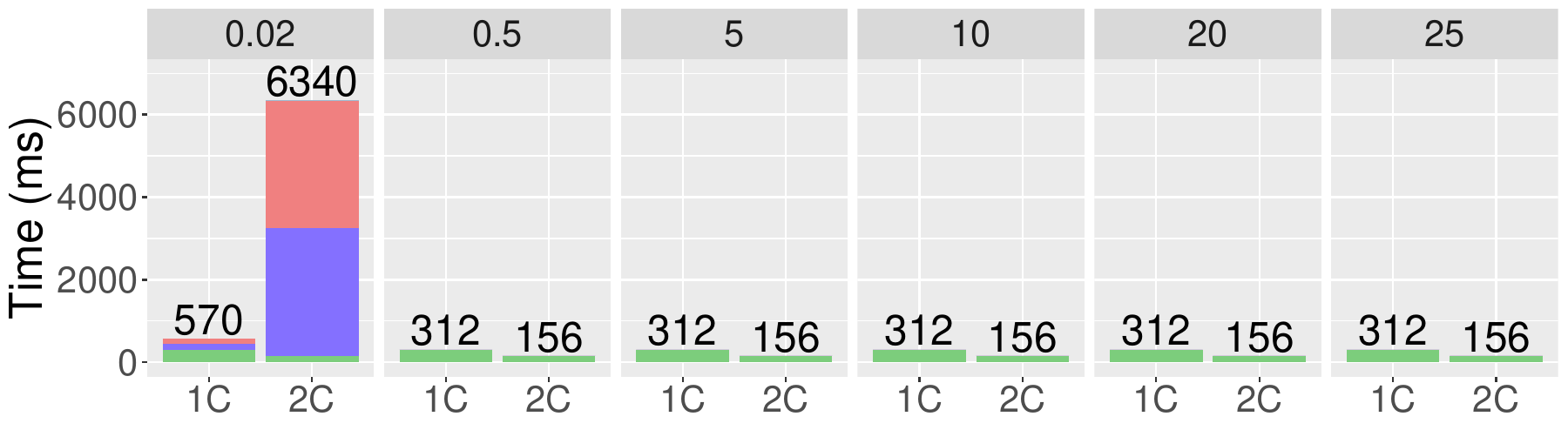}
        \caption{\sysname}
        \label{fig:time-jit-optim}
    \end{subfigure}
    \caption{Execution time comparison.}
    \label{fig:time-evaluation}
\end{figure}

\begin{table*}[ht]
	\tabcolsep=0.25cm
	\newcolumntype{?}{!{\vrule width 1pt}}
	\scriptsize
	\centering
        \caption{Charging time comparison.}
	\begin{tabular}{ c  c  c  c  c  c  c c  c  c  c  c  c}
		\toprule
		~ & \multicolumn{12}{ c }{\textbf{Time (ms)}} \\
            \rowcolor{black!3} 
		~ & \multicolumn{2}{ c }{\emph{0.02 mW}} & \multicolumn{2}{ c }{\emph{0.5 mW} }& \multicolumn{2}{ c }{\emph{5 mW} } & \multicolumn{2}{ c }{\emph{10 mW}} & \multicolumn{2}{ c }{\emph{20 mW}} & \multicolumn{2}{ c }{\emph{25 mW}} \\
            ~ & \emph{1C}& \emph{2C}& \emph{1C}& \emph{2C} & \emph{1C}& \emph{2C} & \emph{1C}& \emph{2C} & \emph{1C}& \emph{2C} & \emph{1C}& \emph{2C} \\
		\midrule
		
		\rowcolor{black!3} 
		\textbf{AdaMICA} & 343172 & 5489716 & 13604 & 148932  & 622   & 2951  & 0 & 494  & 0  & 75  & 0 & 0 \\
		
		\textbf{RockClimb} & 343123 & 5492698 & 14244 &  227851 & 962  & 15784  & 270 & 4744 & 0 & 70  & 0& 35  \\
		
	\rowcolor{black!3} 
		\textbf{\sysname }& 343172 & 5489716 & 4983  & 7596 & 468   & 714 & 0 & 178 & 0 & 89  & 0 & 0 \\
	
		\bottomrule
	\end{tabular}
	
	\label{tab:time-char-evaluation}
\end{table*}

In Figure~\ref{fig:time-evaluation}, we present the comparison of execution time. Multicore intermittent systems start to outperform single-core solutions when input power increases to a certain level, i.e., when the total checkpoint and recovery overhead becomes negligible compared to useful energy consumption. However, RockClimb, running an application on SRAM, cannot benefit from the dual-core parallelism since it needs to execute all the compiler-placed checkpoints. Being power- and energy-aware, \sysname starts to benefit from the parallel execution earlier compared to AdaMICA. For example, even with 0.5mW ambient power, the execution time of our approach consists of only actual computation time, which for the dual-core mode, is 2$\times$ faster than that of the single-core.

In Table~\ref{tab:time-char-evaluation}, we separately compare the charging time for different approaches. With the weakest power presented (0.02mW), the charging time for the application with a 1mF capacitor can reach hundreds and thousands of seconds for all three approaches. However, with increasing input power, the situation changes. For example, with 0.5mW, charging time reduces to tens and hundreds of seconds for AdaMICA and RockClimb and to only several seconds for \sysname. Increasing input power further reduces the charging time, which reaches zero value with maximum power. Note that, depending on the application, even with a 2$\times$ faster dual-core solution, charging time can nullify the parallelization speedup. For example, the 2C \sysname under 20mW input power is twice as fast as 1C with the same parameters, but charging time diminishes the difference to only 27\%. However, greater input power results in greater speedup. Taking charging time into account, \sysname accelerates computation by up to 20$\times$ compared to AdaMICA and up to 30$\times$ compared to RockClimb.

\begin{figure}
    \centering
    \includegraphics[width=1\columnwidth]{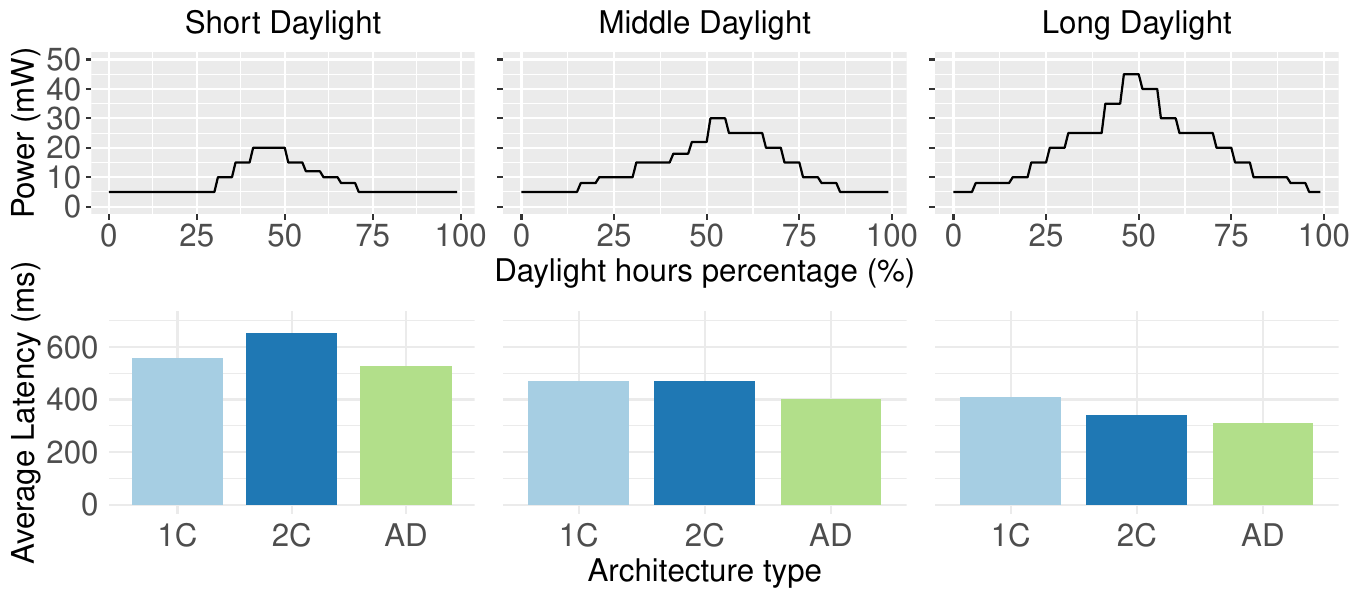}
    \caption{Power-aware adaptation of MAX32666.}
    \label{fig:time-adaptive-max}
\end{figure}

Next, we extend \sysname with the power-aware adaptation (AD), execute 100 applications {\color{black} with 80\% parallelizable code} under three simulated daylight conditions (short, middle, and long), and calculate the average execution latency. {\color{black}This simulation setup reflects the real application scenario (see Section~\ref{sec:real-app}), in which only 80\% of the code is parallelizable. Moreover, the simulation executions take into account the energy overhead of the cameras used in the real application.} As seen in Figure~\ref{fig:time-adaptive-max}, the conditions present different intensities of sunlight power during the day. We consider only the day hours with input power higher than LPMRM power consumption because, with lower power levels, \sysname behaves similarly to the state of the art. With short daylight, we have less intensive power, which prevents the 2C solution from fully benefiting from parallelization. In this scenario, the latency of the 1C solution is improved by 17\%, and the adaptation improves the latency by 6\% more compared to the 2C computation. The middle daylight provides more sunlight power and allows 1C and 2C computations to almost equalize the performance, while the adaptation helps to reduce the latency by 17\% compared to both. The long daylight generates power enough for the dual-core solution to outperform the single-core mode by 19\%. The adaptation in this condition leaves the 1C mode behind by 31\%.

\begin{figure}
    \centering
    \includegraphics[width=1\columnwidth]{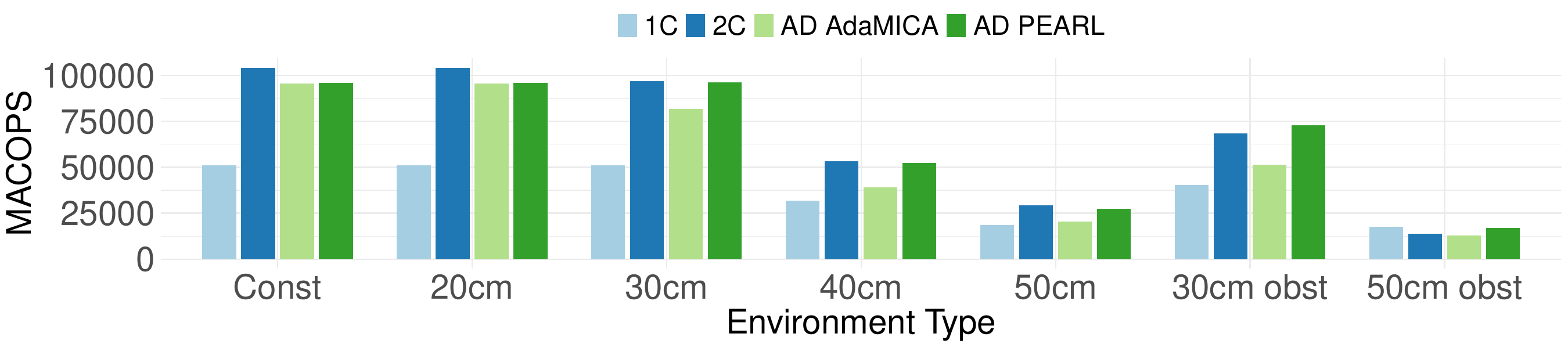}
    \caption{\textcolor{black}{MACOPS comparison.}}
    \label{fig:real-macops}
\end{figure}

\subsection{Testbed Results} 

\label{sec:hardware-results}

We ran the convolution operation under seven different conditions of input power: one with a constant power supply ensuring no power failures; four conditions of harvesting RF energy in different distances between the transmitter and receiver (20, 30, 40, and 50cm); and two energy-harvesting scenarios with different distances and an obstacle periodically appearing (30cm obs and 50cm obs) between the transmitter and receiver. To simulate an obstacle, we put a thin plastic plate (15$\times$20cm) in between every 15 seconds, holding it for 5 seconds. 

Figure~\ref{fig:real-macops} shows MACOPS for all the scenarios. \textcolor{black}{We compare two fixed PEARL implementations, single-core (1C) and dual-core (2C), and two adaptive configurations, PEARL (AD PEARL) and AdaMICA (AD AdaMICA).} 
The constantly powered 2C solution executes twice as fast as the 1C mode performs, while the adaptive solutions perform by almost 94\% better than 1C. The difference between 2C and both AD solutions is due to the decision delay of the adaptation technique. \textcolor{black}{Specifically, the adaptation algorithm of \sysname starts with the single-core mode and waits until the predicted power level reaches the sufficient level, while AdaMICA needs to collect a certain amount of power history to start making adaptation decisions.} Positioning the RF transmitter 20cm away from the receiver does not make any changes in performance compared to the constantly powered environment because the harvested power is strong enough to maintain uninterrupted computation for both 1C and 2C modes. Increasing the distance between the RF transmitter and receiver to 30cm reduces the amount of harvested energy, which causes the interrupts in the 2C execution but keeps the 1C performance unintermittent. The AD modes in this scenario perfectly adjust the architecture to input power. \textcolor{black}{However, compared to AdaMICA, \sysname encounters no power failures, which increases the performance by 15\%.} The 40cm and 50cm distances significantly reduce MACOPS for all the modes, making the 2C mode outperform the 1C mode only by 70\% and 61\%, respectively. \textcolor{black}{Increasing the distance also increases the number of power failures in AdaMICA, allowing \sysname to perform 1.3$\times$ better.} Distracting the RF energy transmission by obstacles during the application execution causes power failures also in \sysname. 
The power failures in the 30cm distance reduce the difference between 1C and 2C MACOPS from 95\% to 70\%, but the adaptive mode in this case outperforms the 1C solution by 83\%. The power failure in the 50cm distance for the 2C mode prevents this mode from outperforming the 1C mode. However, for AD \sysname, a power failure happens during the 1C execution, which is then compensated by faster 2C execution. 

{\color{black}

\begin{table}
  \centering
  \scriptsize
    \caption{\textcolor{black}{ADC average response time and energy.}}
  \begin{tabular}{m{2.5cm}m{2.5cm}m{2.2cm}}
    \toprule
    \rowcolor{black!3}
     \textbf{ADC Sampling Rate (ksps)} & \textbf{Avg. Response Time ($\mu$s)} & \textbf{Energy Overhead ($\mu$J)}\\
    \midrule
    0.1 & 2681 (1072$\times$) &  457  (0.98$\times$) \\
    \rowcolor{black!3}
    1 & 580 (232$\times$)&  4535 (9.7$\times$)  \\
    4 & 160 (64$\times$) & 17675 (38$\times$)  \\
    \rowcolor{black!3}
    7.8 & 39 (16$\times$) &  30477 (65$\times$)\\
    \midrule
    Voltage tracking circuit & 2.5 & 468 \\
    
    \bottomrule
  \end{tabular}
  \label{tab:adc-hml-comp}
\end{table}

\subsubsection{Evaluation of \sysname's Voltage Tracking.} We experimentally assess the response time to the voltage drops and energy overhead of the voltage tracking circuit and ADC-based measurements. We present our results in Table~\ref{tab:adc-hml-comp}. Our measurements concerning several ADC sampling rates show that the response time of the ADC-based voltage tracking can not beat that of the voltage tracking circuit since the maximum ADC sampling rate is 7.8~ksps in MAX32666~\cite{max32666}. At this sampling rate, our circuit detects voltage drops 16$\times$ faster than ADC with the highest sampling rate, consuming 65$\times$ less energy. Besides, the ADC-based approach performs 1072$\times$ slower when it has the same average power consumption of our circuit (0.1~ksps in the table), taking almost 2.6~ms to respond to a voltage drop. These results indicate that the voltage tracking circuit is significantly lightweight in terms of energy consumption and provides a faster response to voltage drops, which is crucial for the correct and efficient operation of \sysname.

\subsection{Real Application Results}

\label{sec:real-app}

As a real-life application scenario, we perform plant disease monitoring that includes data sampling and processing (i.e., CNN inference),  which is aligned with the batteryless remote sensing via image sensor mentioned in Section~\ref{sec:use-cases}. We use a small RGB camera as a default sensor to capture a plant image, whose inference results in approximate accuracy. To increase the accuracy, the second thermal camera can be used with an additional power source. As a default ambient power source, we use a solar panel. As an additional power source, we use the RF transmitter and receiver. We continuously run the application on MAX32666 for 4 minutes for each implementation, AdaMICA, and \sysname, and count the number of approximate and more accurate inferences. Note that we compare \sysname only against AdaMICA since RockClimb is not adaptive and designed only for single-core intermittent systems.

We always use DMA to bring the sensed data to the main memory of the MCU. Thus, active cores always have images to process since bringing data via DMA is much faster than data processing. To emulate the dynamic behavior of environmental power, we change the power source condition every minute in the following order: sunny area, sunny area plus RF, shady area, shady area plus RF. Depending on the strength of ambient power, we exploit two types of parallelism: when only a solar panel is active, we infer two independent RGB images in parallel; with additional RF power, we simultaneously infer one RGB image and one thermal image of the same plant.

\begin{table}
  \centering
  \scriptsize
    \caption{\textcolor{black}{Plant monitoring inference count.}}
  \begin{tabular}{m{1.2cm}m{0.8cm}m{1.3cm}m{0.8cm}m{1.3cm}m{2.3cm}}
    \toprule
    \rowcolor{black!3}
     & \textbf{Sunny} & \textbf{Sunny+RF} & \textbf{Shady} & \textbf{Shady+RF} & \textbf{Overall (appr./prec.)}\\
    \midrule
    AdaMICA & 6 & 5 & 4 & 3 & 10/8 \\
    \rowcolor{black!3}
    \sysname & 12 & 9 & 7 & 5 & 19/14 \\
    
    \bottomrule
  \end{tabular}
  \label{tab:real-life-results}
\end{table}

In Table~\ref{tab:real-life-results}, we compare the number of inferences the application can run in the given time and environmental conditions. Keeping the experimental setup in direct sunlight allows AdaMICA and \sysname to perform in the dual-core mode. However, the power is not strong enough to maintain uninterrupted execution. These power interruptions lead to checkpoints in AdaMICA but cause only sleep mode transfers in \sysname, avoiding all FRAM accesses. The energy and time saved by \sysname allow it to outperform AdaMICA by 2$\times$. Adding the RF power source to the sunny environment allows for activating the thermal camera and conveying more energy for execution. Inferencing from two cameras gives more precise results but increases the inference time and energy almost twice. As a result, with additional RF power, both systems perform fewer but more accurate inferences. 

Moving our test set to a shaded area decreases the performance since AdaMICA and \sysname switch back to the single-core mode. With additional RF power, both systems still cannot transition to the dual-core mode but need to process additional thermal images by a single core. The rightmost column of the table separately presents the overall number of approximate and precise inferences performed, showing that \sysname outperforms AdaMICA by 1.8$\times$. Note that in this experiment, to simplify the memory consistency maintenance of intermittent computing, we disable both cameras and discard all the data belonging to incompletely DMA-transferred images at each power failure and sleep mode transition. We observe that AdaMICA and \sysname spend almost equal time and energy on such wasted data.

\subsection{\sysname Overheads} \sysname has reasonably modest energy and time overheads thanks to its lightweight hardware and software support. In contrast to AdaMICA, \sysname does not utilize power-hungry ADC sampling but three simple comparators providing basic signals (see Section~\ref{sec:hml-impl}). Furthermore, the software part of \sysname requires uncomplicated calculations and very little memory space. 
\sysname properly handles the interrupts triggered by the voltage tracking circuit at reaching $V_{L}$, $V_{M}$, and $V_{H}$. Handling interrupts at $V_{L}$ and $V_{H}$ is a common routine for any just-in-time intermittent system, where $V_{L}$ forces the system to back up the current computational state and $V_{H}$ wakes up the system to restore and continue the interrupted computation. \sysname adds extra actions to the traditional $V_{H}$ interrupt handling: timer initialization, timer read, and power estimation. Additionally, \sysname introduces a $V_{M}$ interrupt handling routine that also initializes and reads the timer value, estimates power, and transitions the system to a sleep mode. 

We isolate and instrument each software routine to accurately capture both time and energy consumption. Execution time was measured using the MAX32666’s on-chip high-resolution timer, which timestamps the start and end of each function. Energy consumption was measured externally using a digital multimeter connected via a high-precision shunt resistor to capture the current drawn during execution, under a constant 3.3V supply. Energy was then computed as the product of current, voltage, and measured duration. Each routine was executed 1000 times, and the values reported in Table~\ref{tab:overheads} reflect the average to reduce the impact of transient variations. This methodology ensures that the reported overheads accurately reflect the cost of PEARL's runtime operations on real hardware.

\begin{table}
  \centering
  \scriptsize
    \caption{\textcolor{black}{\sysname software overheads.}}
  \begin{tabular}{m{2.5cm}m{1.2cm}m{1.3cm}}
    \toprule
    \rowcolor{black!3}
    \textbf{Parameter} & \textbf{Time (\textmu s)} & \textbf{Energy (nJ)}\\
    \midrule
    
    Timer count per ms & 1000 & 1.287\\ 
    \rowcolor{black!3}
    Sleep mode transition & 14 & 0.112\\ 
    $\hat{P}_{amb}^{(1)}$ computation & 0.31 & 3.112 \\
    \rowcolor{black!3}
    $\hat{P}_{amb}^{(2)}$ computation & 0.31 & 3.112 \\
    $\hat{P}$ computation & 0.31 & 3.112 \\
    \rowcolor{black!3}
    Check $Th$ condition & 0.34 & 3.423 \\
      
    \bottomrule
  \end{tabular}
  \label{tab:overheads}
\end{table}

These extra overheads do not hurt the system's performance because they are compensated by a significant decrease in the number of power failures. Furthermore, the timer counter used for power estimation performs simultaneously with application execution and requires only 0.4\textmu A of current. During the entire evaluation process, we observed that the extra \sysname overheads account for just 0.5-1.5\% of the total application time and energy consumption. The memory consumption of the proposed approach is also very modest since it requires no history of power level samples but stores only timer counters and power estimates.

\begin{tcolorbox}[boxsep=2pt,left=2pt,right=2pt,top=0pt,bottom=0pt]
  \paragraph{\textbf{Evaluation Summary.}}
  \sysname avoids all the unnecessary checkpoints, takes benefits from the full utilization of SRAM during computation, and enables the earlier switch to the dual-core mode. These features allow \sysname to outperform the state of the art by 1.7$\times$ to 30$\times$ and to consume 1.5$\times$ to 32$\times$ less energy. The power-aware adaptation of \sysname helps to reduce the execution latency by 17\% to 31\% and improve MACOPS by 29\% to 94\%, compared to the worst of either the 1C or 2C solution proposed by prior works. 
\end{tcolorbox}

\section{Conclusion}
\label{sec:conclusions}

In this paper, we introduced \sysname, a novel systems support that enables efficient intermittent computing on the common off-the-shelf low-power multicore MCU platforms. \sysname connects a three-threshold voltage
tracking circuit and an external fast non-volatile memory to existing ultra-low-power multicore MCUs to make them suitable for efficient multicore intermittent computing. Thanks to these components, \sysname avoids redundant accesses to non-volatile memory and exploits hard-won energy more efficiently. We also presented the \sysname software runtime, which boosts the performance with energy- and power-aware adaptation of the multicore configuration concerning ambient power. Our simulations and real-world experiments showed that \sysname outperforms the state-of-the-art solutions by 1.5 to 32 times in energy efficiency and by 1.7 to 30 times in performance.

\section*{Acknowledgments}

We would like to express our gratitude to our EWSN reviewers for their guidance in refining our final draft. Additionally, we appreciate the anonymous reviewers from EMSOFT 2023, ACM Transactions of Embedded Systems (2024), IEEE Transactions on Emerging Topics in Computing (2024), and EuroSys 2025 for their valuable comments and feedback. 

\begin{wrapfigure}[4]{l}{0.12\columnwidth}
\vspace{-0.62cm}
\includegraphics[width=0.21\columnwidth]{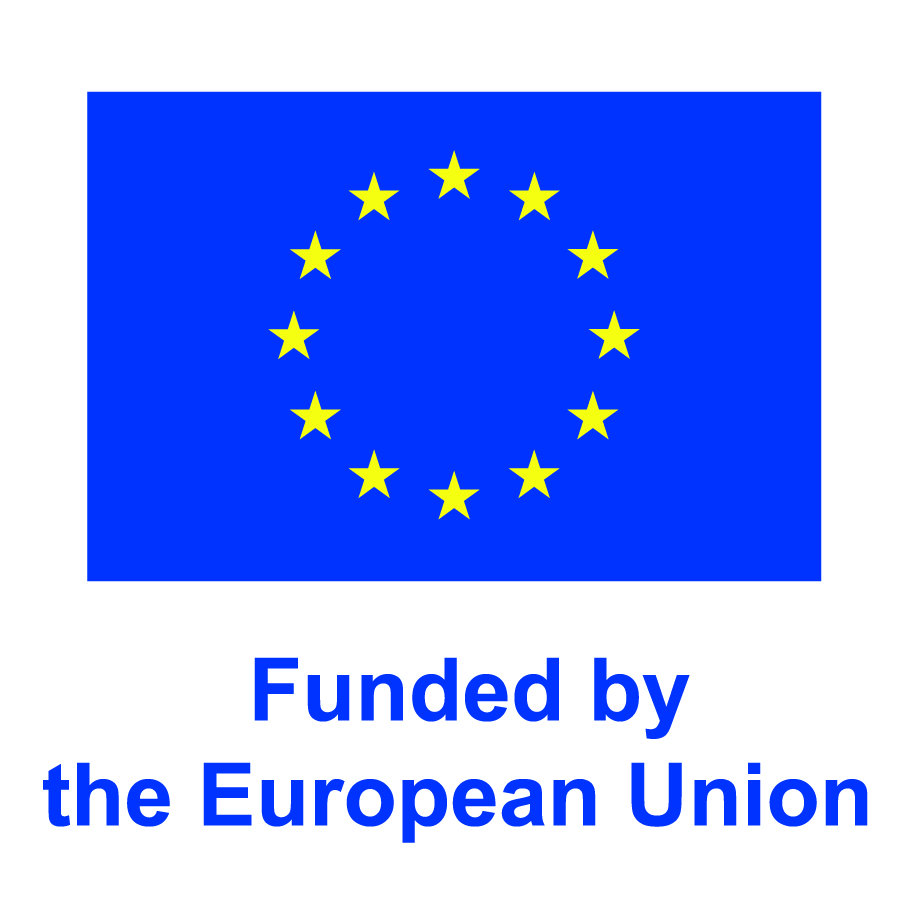}
\end{wrapfigure}%
\noindent Funded by the European Union (HE - Crosscon - GA  101070537). Views and opinions expressed are however those of the author(s) only and do not necessarily reflect those of the European Union or the European Commission. Neither the European Union nor the granting authority can be held responsible for them.


\pagebreak

\bibliographystyle{plain}
\bibliography{references}

\end{document}